\documentstyle[aps,preprint,prc,12pt]{revtex}

\begin{document}

\title{
Second Order Phase Transitions : From Infinite to Finite Systems
}

\author{
P. Finocchiaro$^{1}$, M. Belkacem$^{1}$, T. Kubo$^{1}$\footnote{present
address: Universit\"{a}t T\"{u}bingen, Institut f\"{u}r Theoretische Physik,
D-71076 T\"{u}bingen, Germany},
V. Latora$^{1,2}$, and A. Bonasera$^{1}$
}

\address{
$^{1}$ INFN - Laboratorio Nazionale del Sud \\
Viale Andrea Doria (ang. Via S. Sofia), 95123 Catania, Italy \\
$^{2}$ Dipartimento di Fisica, Universit\`{a} di Catania \\
57, Corso Italia, 95129 Catania, Italy
}

\date{ \today }
\maketitle


\begin{abstract}
We investigate the Equation of State (EOS) of classical systems having 300
and 512 particles confined in a box with periodic boundary conditions.
We show that such a system, independently on the number of particles
investigated, has
a critical density of about 1/3 the ground state density and a critical
temperature of about $2.5~ MeV$.  The mass distribution
at the critical point exhibits a power law with $\tau = 2.23$.
Making use of the grand partition function of Fisher's droplet
model,  we obtain an analytical EOS around the
critical point in good agreement with the one extracted from the numerical
simulations.

\end{abstract}
{
\vskip 2\baselineskip
}
\newpage

\section {Introduction}

Recently, several experiments in proton-nucleus and nucleus-nucleus
reactions have revealed the creation of fragment size distributions exhibiting
power laws, expected according to the droplet model of Fisher, for fragment
formation near the critical point of a liquid-gas phase transition \cite{mahi}.
This observation raised the question whether critical phenomena, which strictly
speaking can only occur in the thermodynamic limit, can be explored within
the context of heavy-ion collisions, where only a few hundreds of nucleons
are involved \cite{balescu}.

In  previous works \cite{belkacem}, by studying the expansion
of a classical finite system within the framework of Classical Molecular
Dynamics (CMD),  evidence has been found for the occurrence of a critical
behaviour of our system revealed through a study of fragment size
distributions,
scaled factorial moments and moments of size distributions. Such a critical
behaviour is connected to a liquid-gas phase transition by the use of Fisher's
droplet model \cite{fisher} and Campi analysis \cite{campi}. In the present
work, using the same two-body interaction \cite{pan90},
we calculate by means of the virial
theorem \cite{balescu,pratt},
 the equation of state of a classical system confined in a cubic box
with constant density and periodic boundary conditions. This equation of state
has a critical temperature of about 2.5 $MeV$ and a critical density of about
one third the ground state density. Trajectories in the ($\rho$, $T$)
plane of expanding finite systems indeed go through this critical point
\cite{belkacem}. It is also shown that at the critical point, the fragment
size distribution exhibits a power law with $\tau = 2.23$, consistent with the
droplet model of Fisher \cite{belkacem,fisher}, and scaling laws of critical
exponents for a liquid-gas phase transition.

We calculate in Section II, within the framework of CMD \cite{belkacem,pan90},
 the equation of state
of the system in a cubic box with cyclic boundary conditions. Section
III contains the study of fragment size distributions at the critical point and
around it. In Section IV, we study the critical behaviour of the moments of
size distributions using the grand partition function of the Fisher's droplet
model and extract the equation of state of this model in the vicinity of the
critical point, which is found in good agreement with the one calculated
for the infinite system. Finally, we give our summary and conclusions in
Section V.

\section {Equation of State}

In the following, making use of the virial theorem \cite{balescu,pratt},
we study the
equation of state of classical systems with 300 particles (150 protons +
150 neutrons) and 512 particles (256 protons + 256 neutrons) confined
in a cubic box with periodic boundary conditions.
The particles of the system move under the influence
of a two-body potential $V$ given by \cite{pan90} :
\begin{eqnarray}
V_{np}(r) &=& V_{r}\left[exp(-\mu_{r}r)/{r} - exp(-\mu_{r}r_{c})/{r_c}\right]
\nonumber \\
          & & \mbox{}- V_{a}\left[exp(-\mu_{a}r)/{r} -
exp(-\mu_{a}r_{a})/{r_a}\right]
\nonumber \\
V_{nn}(r) &=& V_{pp}(r) = V_{0}\left[exp(-\mu_{0}r)/{r} -
exp(-\mu_{0}r_{c})/{r_c}\right]
\label{eq1}
\end{eqnarray}
 ${r_c}=5.4~ fm $ is a cutoff radius.
$V_{np}$ is the potential acting between a neutron and a proton while $V_{nn}$
is the potential acting between two identical nucleons.
The first potential is attractive at large $r$ and repulsive at small $r$,
while the latter is purely repulsive so no
bound state of identical nucleons can exist.
The various parameters entering eq.(1) are defined with their respective
values in Ref. \cite{pan90}.
The classical
Hamilton's equations of motion are solved using the Taylor
method at the order O[$(\delta t)^2$]  where $\delta t$ is the integration
time step \cite{computer}.  Energy and momentum are well conserved because
we use a very small time step $\delta t$ varying in time according to the
maximum acceleration of the particles.

Two different methods were used for initialization and thermalization of the
system. In the first one, the particles are initially distributed randomly
in a cubic box with a small size (the initial density is about $0.32~ fm^{-3}$)
and they are given the same velocity modulus which is quite large. The
system is then cooled down to a predefined average kinetic energy by
rescaling the velocities at each time iteration by a coefficient close to
unity. After the density is decreased slowly to the desired
density (at each time iteration the size of the box is slowly increased).
During the cooling process, the system also thermalizes. Nevertheless, after
the desired density and temperature have been reached, we let
it evolve for additional time to allow a complete thermalization.
In the second method, the particles are initially distributed on a cubic
lattice
such as the nearest neighbors of a proton are all neutrons and vice versa, and
the system is  excited at an initial temperature giving to its particles a
maxwellian velocity
distribution by means of a Metropolis sampling \cite{computer}. The system then
evolves according to Hamilton's equations of motion and reaches equilibrium
after few thousands of iterations. With the first method, the system reaches
equilibrium faster than with the second one.

In Fig. 1 we have plotted the time evolution of the kinetic energy
(upper part), the potential energy (lower part, solid line) and total energy
(lower part, dashed line) computed using the second method for thermalization
for the system with 512 particles. The density of the system is fixed at
$\rho =0.16~ fm^{-3}$ and the initial temperature is set at $T=1.66~ MeV$.
We see that the total energy is well conserved during all the time evolution
($2.5\%$ in $100~ fm/c$). In this case the system thermalizes very quickly
in the first $20~ fm/c$, then the kinetic energy stabilizes around 1.0 $MeV$
($T = 0.7~ MeV$) with small
fluctuations due to the interplay between kinetic and potential energy.
In Fig. 2 we plot the same quantities at a density of $0.04~ fm^{-3}$
and the same initial temperature $T=1.66~ MeV$. In this case we see that the
thermalization process takes a longer time and the kinetic energy stabilizes
around 4. $MeV$.

In the following, the calculations are done using the first method of
initialization and thermalization for the system with 300 particles
(150 protons + 150 neutrons). The results obtained with the second method for
the system with 512 particles (256 protons + 256 neutrons)
are totally consistent with those presented here and give the same equation of
state with the same critical point.

After the system has thermalized, we calculate the equation of state of our
system by means of the virial theorem. The pressure of a system of classical
particles interacting through two-body forces is given by
\cite{balescu,dorso,pratt}
\begin{eqnarray}
P(\rho, T) &=& \rho T + \frac{1}{6V} \int d{\bf r}_{1} d{\bf r}_{2}
\left| {\bf r}_{1} - {\bf r}_{2} \right|
F_{12}({\bf r}_{1} - {\bf r}_{2}) n_{2}({\bf r}_{1},{\bf r}_{2})
\nonumber \\
&=& \frac{2}{3V} \left< \sum_{i} E_{i} \right> + \frac{1}{3V}
\left< \sum_{i<j} \left|{\bf r}_{i} - {\bf r}_{j}\right| \cdot F_{ij}
({\bf r}_{i} - {\bf r}_{j}) \right>
\label{eq2}
\end{eqnarray}
where $V$ is the volume of the box, $\rho = A/V$ the density,
$F_{ij} = -dV_{ij}/dr$ is the force acting between particles $i$ and $j$, and
$n_{2}({\bf r}_{1},{\bf r}_{2})$ is the pair distribution function. $E_{i}$ is
the kinetic energy of particle $i$, and the brackets $< \cdot >$ denote the
time
average. In the above equation, the Boltzmann constant $k$ is set equal to 1.
To calculate the pressure, we let the system evolve after the
thermalization for additional time and take the average over this time
evolution. The density of the system is fixed, while the temperature is given
in the equilibrium phase by 2/3 of the kinetic energy.

In Fig. 3 we have plotted the pressure $P$ versus the final
temperature of the system $T$ for several values of the density $\rho$
ranging from
0.02 to 0.24 $fm^{-3}$. We see from the figure that for all the values of the
density considered, the pressure $P$ shows a linear behaviour versus
temperature,
\begin{equation}
P(\rho, T) = a(\rho)T + b(\rho)
\label{eq3}
\end{equation}
Note that the temperatures considered here are that of the thermalized system
(at equilibrium).
The values of the fitting parameters $a(\rho)$ and $b(\rho)$ are plotted in
the upper and lower parts of
Fig. 4, respectively,  versus the density $\rho$ (solid dots). At this point,
to determine
the equation of state, we fitted the parameters $a(\rho)$
and $b(\rho)$ with polynomials in $\rho$ with the minimum number of fitting
parameters and the following constraints: {\it i}) All the isotherms have a
zero pressure at zero density $P(\rho = 0,T) = 0$, which means
$a(0) = b(0) = 0$; {\it ii}) The isotherm $T = 0~ MeV$ has a zero slope at
zero density
 $\frac{\partial P}{\partial\rho}\big\vert_{T=0,\rho=0} = 0$, which
means $\frac{\partial b}{\partial\rho}\big\vert_{\rho=0} = 0$. The resulting
polynomial fits give:
\begin{eqnarray}
P(\rho, T) &=& (a_{1}\rho + a_{2}\rho^{2})T + b_{2}\rho^{2} + b_{3}\rho^{3}
\nonumber \\
&=& (0.96\rho + 7.13\rho^{2})T - 87.0\rho^{2} + 646.0\rho^{3}
\label{eq4}
\end{eqnarray}
The fits to $a(\rho)$ and $b(\rho)$ are shown by solid lines in Fig. 4.
Note that the parameter $a_{1}$ is almost equal to 1 which means that our
equation of state is in agreement with that of a classical gas at high
temperature
and low density $\rho \rightarrow 0, T \rightarrow \infty \Longrightarrow
P \approx \rho T$. Turning off the two-body potential, the law of ideal gases
is respected within less than $1\%$. We
have plotted in Fig. 5 the pressure $P$, eq.(4),
 versus density for several values
of the temperature $T$. The dashed line in the figure delimits the isothermal
spinodal region given by
\begin{equation}
\frac{\partial P}{\partial \rho}\big\vert_{T} < 0
\label{eq5}
\end{equation}

The obtained equation of state eq.(\ref{eq4}) has a critical point
\begin{equation}
T_{c} = 2.53~ MeV~~~;~~~~\rho_{c} = 0.04~ fm^{-3}~~~;~~~~P_{c} = 0.029~
MeV fm^{-3}
\label{eq5_1}
\end{equation}
Note that the ratio $P_{c}/T_{c}\rho_{c} = 0.32$ is in good agreement with the
experimental values for real gases which give for example 0.287 for $CO_{2}$,
0.290 for $Xe$ and 0.305 for $^4He$. We have also plotted in Fig. 6 the
isothermal spinodal line (eq.(\ref{eq5})) in the plane ($\rho/\rho_{c}$,
$T/T_{c}$) together with the experimental values for eight different substances
in the gas-liquid coexistence region \cite{huang}. The critical data for all
these substances are quite varied (see Ref. \cite{huang}), but the reduced data
points fall on a universal curve. The right branch refers to the liquid phase,
and the left branch to the gas phase, and both come together at the critical
point. Our coexistence curve falls also, with some small differences, on the
same universal curve. The dashed line shows Guggenheim's fit \cite{huang}.

As already stated, using the second method of
initialization and thermalization for the system with 512 particles we
obtained the same equation of state eq.(\ref{eq4})
with less that $5\%$ difference with the first method using a system with
300 particles. This allows us to conclude that the equation of state of our
system does not depend on the number of particles we have used in the box,
 and on
the way how the system has reached equilibrium.

\section{Mass Distributions}

In this section, our aim is to generate the mass distributions at the critical
point and around it to see whether one gets a power law at the critical point
as predicted by the droplet model of Fisher \cite{fisher}, and as observed in
CMD for an expanding classical system \cite{belkacem}. To do so, we generate
many events near the critical temperature, and  at the critical
density $\rho_{c} = 0.04~ fm^{-3}$. Unfortunately our algorithm
of cluster recognition which is of the percolation-type, is not able to
identify the fragments inside the box, even at such small density
\cite{belkacem1}. In this
algorithm, the fragments are defined as follows. It is possible to go from any
particle $i$ to another particle $j$ belonging to the same fragment by
successive interparticle jumps of a prescribed distance $d$ or less (in these
calculations, $d$ is set equal to the range of the two-body potentials
$d = r_{c} = 5.4~ fm$), but any path to a particle $k$ belonging to another
fragment contains one or more jumps of distance greater than $d$. This kind
of cluster recognition algorithms is not indicated for identifying fragments
confined in a box because the results are very dependent on the choice of the
distance $d$. However, for expanding systems with fragments flying away from
each other, these algorithms work well, particularly by following
the fragments in time. To overcome this inherent problem, after the system has
thermalized and the fragments are formed inside the box, we take off the box
and let the system expand and the fragments fly away for enough time to
allow a good identification of the fragments.

Because this procedure is
very CPU-time consuming, we have generated only 100 events of the system with
216 particles (108 protons + 108 neutrons) at each temperature
and calculated the mass distributions from these events. The results of
these calculations are drawn in Fig. 7, where we have plotted the mass
distributions obtained from the expansion of system prepared at the critical
density $\rho_{c} = 0.04~ fm^{-3}$ and temperatures (at thermalization)
$T= 2.0~ MeV$, $T_{c}= 2.54~ MeV$ and $T= 3.92~ MeV$. At the critical
temperature,
apart for large masses where one observes large fluctuations due to the
limited number of events generated, one observes a power law,
consistent with the droplet model of Fisher, with a power $\tau = -2.23$, the
same as the one found for the expanding finite system passing during its
expansion through the critical point. Below the critical temperature, one
observes a mass distribution with a "U" shape characteristic to undercritical
events where one sees some remaining of the initial system, and above the
critical temperature one observes a rapidly decreasing mass
distribution with an exponential shape characteristic to highly excited systems
going to complete vaporization.

\section{Fisher's Droplet Model and the Equation of State}

Fisher's droplet model has successfully been applied to describe the
fragmentation of excited classical systems which expand and, depending on the
excitation energy show different dynamical evolutions, from evaporation like
processes (small excitation energy) to multifragmentation and complete
vaporization of the
system (for large excitation energies) \cite{belkacem}.

In the following, using the grand partition function Fisher has derived for a
large system of individual particles in which clusters of two, three and more
particles are bound together by the attractive forces, and in mutual
statistical equilibrium with the remaining parts of the system and the
surrounding gas \cite{fisher1}. In this model, the grand partition function
\begin{equation}
\Xi(z,V,\beta) = \sum_{N=0}^{\infty} z^{N} Q_{N}(V,\beta)
\nonumber
\end{equation}
where $z = e^{\beta\mu}$ is the fugacity, $\beta = 1/T$ and $Q_{N}(V,\beta)$
is the canonical partition function, is given by
\begin{equation}
ln \Xi(\mu,V,T) = \sum_{A=1}^{\infty} Y(A)
\label{eq6}
\end{equation}
where $Y(A)$ is the probability to find a cluster of size $A$ in the system
(cluster size distribution) and is given by \cite{fisher1}
\begin{eqnarray}
Y(A) &=& Y_{0} A^{-\tau} exp\left[\beta\mu A - \beta b A^{2/3}\right]
\nonumber \\
 &=& Y_{0} A^{-\tau} B^{A^{2/3}} S^{A}
\label{eq7}
\end{eqnarray}
where $\mu = \mu_{g} - \mu_{l}$ is the difference between the chemical
potentials of the gas and liquid phases. $b$ denotes the surface contribution
to the Gibbs free energy and is proportional to the surface tension. The term
$A^{-\tau}$ has been introduced by Fisher to take into account the fact that
the droplet surface closes on itself. The parameter $\tau$ is related to some
critical exponents through scaling laws of critical phenomena and thus has a
constant value in the range $2 < \tau < 2.5$ \cite{fisher,fisher1}.
Knowing the grand partition function, the
equation of state of the system is given by
\begin{eqnarray}
P &=& \frac{T}{V} ln \Xi(\mu,V,T) = \frac{T}{V} \sum_{A=1}^{\infty} Y(A)
\nonumber \\
<N> &=& z\frac{\partial}{\partial z} ln \Xi(\mu,V,T)
= \sum_{A=1}^{\infty} A Y(A)
\label{eq8}
\end{eqnarray}
and the isothermal compressibility is given by,
\begin{eqnarray}
\chi_{T} &=& \frac{VT}{<N>^{2}} \frac{\partial^{2}}{\partial \mu^{2}}
ln \Xi(\mu,V,T)
\nonumber \\
&=& \frac{V}{T<N>^{2}} \sum_{A=1}^{\infty} A^{2} Y(A)
\label{eq9}
\end{eqnarray}
Defining the moments of the cluster size distributions $Y(A)$ by
\begin{equation}
M_{k} = \sum_{A=1}^{\infty} A^{k} Y(A)
\label{eq10}
\end{equation}
gives for the previous quantities
\begin{eqnarray}
P &=& \frac{T}{V} M_{0}
\nonumber \\
<N> &=& M_{1}~~\Longrightarrow~~ \rho = \frac{<N>}{V} = \frac{M_{1}}{V}
\label{eq11}
\end{eqnarray}
and
\begin{equation}
\chi_{T} = \frac{1}{VT \rho^{2}} M_{2}
\label{eq12}
\end{equation}
{}From eqs.(\ref{eq11}), one obtains the equation of state in terms of the
zeroth and first moments of the cluster size distribution
\begin{equation}
P = T\rho M_{0}/M_{1}
\label{eq13}
\end{equation}

More generally, the moments of the cluster size distributions are related
to Fisher's grand partition function by:
\begin{equation}
M_{k} = \left(T \frac{\partial}{\partial \mu}\right)^{k}
ln \Xi(\mu,V,T)
\label{eq14}
\end{equation}

The equation of state eq.(\ref{eq13}) is useless because of the complexity
in estimating in a model independent way the cluster size distributions $Y(A)$
(eq.(\ref{eq7})) and hence the moments $M_{k}$. However, an estimate of the
moments in the vicinity of the critical point is possible by making a few
approximations. In the formula for the cluster size distributions
eq.(\ref{eq7}), for temperatures equal or larger than the critical temperature
$T \geq T_{c}$, the surface term $b$ vanishes because it is proportional to
the surface tension which is equal to 0 for $T \geq T_{c}$ and the difference
of chemical potentials $\mu$ is negative
\begin{equation}
Y(A) = Y_{0} A^{-\tau} exp(\beta \mu A)~~~~ T \geq T_{c}
\label{eq15}
\end{equation}
We can now use the well known formula by Robinson derived for the
Bose-Einstein integral \cite{robinson} :
\begin{equation}
\sum_{A=1}^{\infty} e^{-A\alpha} A^{-\sigma} = \Gamma(1-\sigma)
\alpha^{\sigma - 1} + \sum_{n=0}^{\infty} \frac{(-1)^{n}}{n!}
\zeta(\sigma - n)\alpha^{n}
\label{eq16}
\end{equation}
where $\alpha > 0$.

Using this formula, the moments of the cluster size distribution become for
$T \geq T_{c}$ :
\begin{equation}
M_{k} = Y_{0} \left[\Gamma(1 + k - \tau) |\beta \mu|^{-(1+k-\tau)}
+ \sum_{n=0}^{\infty} \frac{(-1)^{n}}{n!}
\zeta(\tau - k - n)|\beta \mu|^{n} \right]
\label{eq17}
\end{equation}
The second term in the rhs of the above equation is regular and hence, while
the zeroth and first moments are always finite because $\tau$ is limited
between 2 and 2.5, the higher moments diverge as
\begin{equation}
M_{k} \propto |\beta \mu|^{-(1+k-\tau)}
\label{eq18}
\end{equation}
By assuming $\mu = (T - T_{c})^{\nu}$, one recovers the formula Campi has
derived for the moments at the critical point \cite{campi}
\begin{equation}
M_{k} \propto |T - T_{c}|^{-\nu(1+k-\tau)}
\label{eq19}
\end{equation}

In the vicinity of the critical point, the difference of chemical potentials
$\mu$ tends to zero and one can consider the lowest orders only in the
expansion of the moments $M_{k}$ in terms of powers of $\mu$, eq.(\ref{eq17}),
which gives for the pressure $P$ and density $\rho$ (assuming $\tau = 2.33$):
\begin{eqnarray}
P &=& q_{0}T\left[3.072|\beta\mu|^{4/3} + 1.417 - 3.631|\beta\mu| + ...\right]
\nonumber \\
\rho &=& q_{0}\left[-4.086|\beta\mu|^{1/3} + 3.631 + 0.966|\beta\mu| +
...\right]
\label{eq20}
\end{eqnarray}
where $q_{0} = Y_{0}/V$. At the critical point, $\mu = 0$ and one gets,
\begin{eqnarray}
P_{c} &=& 1.417 q_{0} T_{c}
\nonumber \\
\rho_{c} &=& 3.631 q_{0}
\label{eq21}
\end{eqnarray}
Subtracting eqs.(\ref{eq20}) and (\ref{eq21}) gives,
\begin{eqnarray}
P - P_{c} &=& q_{0}T\left[3.072|\beta\mu|^{4/3} - 3.631|\beta\mu| +
...\right] + 1.417 q_{0} (T - T_{c})
\nonumber \\
\rho - \rho_{c} &=& q_{0}\left[-4.086|\beta\mu|^{1/3} +
0.966|\beta\mu| + ...\right]
\label{eq22}
\end{eqnarray}
Combining the above equations and taking the lowest orders in terms of powers
of
$\mu$, we get the following equation of state, valid in the vicinity of the
critical point
\begin{equation}
P - P_{c} = 1.417 q_{0} (T - T_{c}) + \frac{0.0532}{q_{0}^{2}} T (\rho -
\rho_{c})^{3} + \frac{0.011}{q_{0}^{3}} T (\rho - \rho_{c})^{4}
\label{eq23}
\end{equation}

Using the numerical values of the critical temperature, density and pressure,
obtained in section II for the system in a box with periodic
boundary conditions (eq.(\ref{eq5_1})), we have plotted in Fig. 8, the
critical isotherm of the above equation of state in the vicinity of the
critical
point (dashed line). In the same figure, we have also plotted the critical
isotherm of the equation of state obtained in section II eq.(\ref{eq4}) (solid
line). The
critical point is indicated in the figure by the open circle.
 We see from the
figure that the two curves go on each other in the vicinity of the critical
point, while they start to deviate going away from the
critical region.

{}From eq.(\ref{eq14}), we see that the isothermal compressibility $\chi_{T}$
is
related to the second moment $M_{2}$. At the critical point, the isothermal
compressibility becomes infinite, and so the second moment diverges, which
means that the fluctuations of cluster sizes expressed by the second moment
$M_{2}$ become very large at the critical point. Defining the relative
variance $\gamma_{2}$, introduced by Campi to have more insight in the shape
of the cluster size distributions and to indicate the critical behaviour
\cite{campi},
\begin{equation}
\gamma_{2} = \frac{M_{2} M_{0}}{M_{1}^{2}}
\end{equation}
one expects that this quantity diverges at the critical point. Of course,
if one considers a finite system, the moments remain finite, and instead of
a divergence, the relative variance $\gamma_{2}$ shows a peak around the
critical point. To study finite size effects on the relative variance
$\gamma_{2}$, we calculate the moments of cluster size distributions $Y(A)$
using for the parameters $Y_{0}$, $\tau$, $B$ and $S$ appearing in
eq.(\ref{eq7}) the values obtained
by fitting the fragment mass distributions obtained in Molecular Dynamics
simulations of an expanding system with different initial excitation energies
(temperatures) \cite{belkacem}. In Fig. 9 we have plotted the relative variance
$\gamma_{2}$ versus the initial temperature (excitation energy) for different
values of the total size of the
system $A_{tot}$. One can see the signature of the critical point for
the initial temperature $T = 5~ MeV$ where one exactly obtains a power law
in the fragment size distribution. One also sees that the reduced variance
diverges more and more sharply as $A_{tot}$ increases. It is also interesting
to compare the critical ratio $P_{c}/\rho_{c}T_{c} = M_{0}/M_{1}|_{T_{c}}$
calculated at the point where one observes the peak for $\gamma_{2}$, with
other systems. For $A_{tot} = 100$ and $A_{tot} = 10^{8}$, one gets
$M_{0}/M_{1} = 0.299$ and 0.298, respectively. Our numerical simulation of
the system in a box gives 0.317, and Van der Waals equation gives 0.375. Real
gases give 0.287 for $CO_{2}$, 0.290 for $Xe$ and 0.305 for $^4He$
\cite{huang}.

\section{Conclusions}

We have calculated by means of the virial theorem the equation of state of an
infinite classical system in a cubic geometry with periodic boundary conditions
and constant density. This equation of state has a critical temperature of
2.53 $MeV$ and a critical density of 0.036 $fm^{-3}$, about 1/3 the ground
state density. We note that by using different numbers of particles in the box
and different initialization and thermalization procedures, we obtain the same
EOS with the same critical point. Another point is the decreasing of the
critical temperature with the number of particles of the system. Several
investigations of finite size effects on the critical temperature
using Skyrme-type interactions \cite{mekjian} have concluded that the critical
temperature goes down when decreasing the number of particles of the system. In
our present work, it appears that this is not true and we have obtained for an
infinite system (with periodic boundary conditions) the same critical
temperature as the one observed in a previous work where we have studied, using
the same two-body interaction the expansion of a finite system with 100
particles, passing during its evolution through the critical point.
Furthermore,
we have obtained at the critical point a fragment size distribution showing a
power law with $\tau=2.23$. The same power law was also observed for the
expanding finite system for the critical evolution.

Finally, by making use of the grand partition function of the droplet model of
Fisher, we have studied the critical behaviour of the moments of fragment
size distributions and found the same behaviour as the one calculated by Campi.
We have also extracted the equation of state of this model around the critical
point shown that it is in good agreement with that calculated for the
infinite classical system.

\centerline{\bf Acknowledgements}

The authors acknowledge S. Ayik, M. Di Toro and V. N. Kondratyev for
stimulating discussions. One of us (T. K) thanks the theory group at INFN-LNS
Catania for partial support and hospitality where the major part of this work
was carried out.

\newpage


\begin{figure}
\caption{Kinetic energy (upper part), potential energy (lower part, solid line)
and total energy (lower part, dashed line) are plotted versus time for the
system with 512 particles using the second method for thermalization (see
text). The density is fixed at $\rho = 0.16~ fm^{-3}$
and the system is initialized at a temperature $T = 1.66~ MeV$.}
\end{figure}

\begin{figure}
\caption{Same as Figure 1 but the density of the system is fixed at $\rho =
0.04~ fm^{-3}$.}
\end{figure}

\begin{figure}
\caption{The pressure $P$ of the system, calculated using the virial theorem,
is plotted versus temperature $T$ for several values of the density ranging
from 0.03 to 0.24 $fm^{-3}$. The points indicate the results of the numerical
simulations, and the solid lines the linear fits. The temperatures considered
are the ones at equilibrium.}
\end{figure}

\begin{figure}
\caption{The values of the fitting parameters $a(\rho)$ (upper part) and
$b(\rho)$ (lower part) are plotted versus density $\rho$ (solid points). The
solid lines show the polynomial fits to these values (see text).}
\end{figure}

\begin{figure}
\caption{The pressure $P$, calculated using the obtained equation of state
eq.(4), is plotted versus density $\rho$ for several values of temperature $T$.
The dashed line indicates the isothermal spinodal line of the EOS. The critical
isotherm is indicated by the small-dashes line.}
\end{figure}

\begin{figure}
\caption{Reduced temperature versus reduced density in the gas-liquid
coexistence region for our system (solid line) and for eight different
substances; $Ne$, $A$, $Kr$, $Xe$, $N_{2}$, $O_{2}$, $CO$ and $CH_{4}$ (open
circles). The dashed line shows Guggenheim's fit.}
\end{figure}

\begin{figure}
\caption{Mass distributions obtained for the fragmentation of the system with
216 particles with tree different temperatures (at equilibrium) $T=2.0~ MeV$
(upper part), $T_{c} = 2.54~ MeV$ (central part) and $T = 3.92~ MeV$ (lower
part).}
\end{figure}

\begin{figure}
\caption{Pressure versus density. The solid line shows the critical
isotherm of the equation of state obtained in Section II by numerical
simulations of a classical system, and the dashed line the critical isotherm of
the analytical EOS obtained in Section IV using Fisher's droplet model. The
open circle indicates the critical point.}
\end{figure}

\begin{figure}
\caption{The reduced variance $\gamma_{2}$ is plotted versus the initial
temperature for the expansion of a finite system (see text) for
several values of the size of the system $A_{tot}$. }
\end{figure}


\newpage

\begin{thebibliography}{99}

\bibitem{mahi} J. E. Finn et al., Phys. Rev. Lett. 49 (1982) 1321;
Phys. Lett. B 118 (1982) 458;

H. H. Gutbrod, A. I. Warwick and H. Wieman, Nucl. Phys. A 387 (1982) 177c;

M. Mahi, A. T. Bujak, D. D. Carmony, Y. H. Chung, L. J. Gutay, A. S. Hirsch,
G. L. Paderewski, N. T. Porile, T. C. Sangster, R. P. Scharenberg and B. C.
Stringfellow, Phys. Rev. Lett. 60 (1988) 1936;

J. B. Elliot, M. L. Gilkes, J. A. Hauger, A. S. Hirsch, E. Hjort,
N. T. Porile, R. P. Scharenberg, B. K. Srivastava, M. L. Tincknell and P. G.
Warren, Phys. Rev. C 49 (1994) 3185;

M. L. Gilkes et al., Phys. Rev. Lett. 73 (1994) 1590.

\bibitem{balescu} R. Balescu, Equilibrium and Nonequilibrium Statistical
Mechanics (Krieger Publishing Company, Malabar, Florida, 1991).

\bibitem{belkacem} V. Latora, M. Belkacem and A. Bonasera, Phys. Rev. Lett.
73 (1994) 1765;

M. Belkacem, V. Latora and A. Bonasera, Phys. Rev. C 52 (1995) 271.


\bibitem{fisher} M. E. Fisher, Rep. Prog. Phys. 30 (1967) 615;
Proc. International School of Physics, Enrico Fermi Course LI, Critical
Phenomena, ed. M. S. Green (Academic, New York, 1971).

\bibitem{campi} X. Campi, J. of Phys. A 19 (1986) L917;
Phys. Lett. B 208 (1988) 351; J. de Phys. 50 (1989) 183.

\bibitem{pan90} R. J. Lenk, T. J. Schlagel and V. R. Pandharipande,
Phys. Rev. C 42 (1990) 372.

\bibitem{dorso} C. O. Dorso and J. Randrup, Phys. Lett. B 232 (1989) 29.

\bibitem{pratt} S. Pratt, C. Montoya and F. Ronning, Phys. Lett B 349 (1995)
261.

\bibitem{computer} S. E. Koonin and D. C. Meredith, Computational Physics
(Addison Wesley, California, 1990).

\bibitem{huang} K. Huang, Statistical Mechanics (John Wiley $\&$ Sons,
New York, 1987).

\bibitem{belkacem1} A. Bonasera, F. Gulminelli and J. Molitoris, Phys. Rep.
243 (1994) 1;

M. Belkacem, V. Latora and A. Bonasera, Phys. Lett. B 326 (1994) 21.

\bibitem{fisher1} M. E. Fisher, Physics 3 (1967) 255.

\bibitem{robinson} J. E. Robinson, Phys. Rev. 83 (1951) 678.

\bibitem{mekjian} H. R. Jaqaman, A. Z. Mekjian and L. Zamick, Phys. Rev. C
27 (1983) 2782; Phys. Rev. C 29 (1984) 2067.

\end{thebibliography}
\end{document}